\documentstyle[12pt,axodraw]{article}
\newcommand{\prepr}[1]
{\begin{flushright} {\bf #1} \end{flushright} \vskip 1.5cm}
\newcommand{\titul}[1] {\begin{center}{\large\bf #1 } \end{center}\vskip 1.cm}
\newcommand{\autor}[1] {\begin {center} {\large \lineskip .5em #1 }
                        \end   {center} }
\newcommand{\lugar}[1] {\begin{center} {\it #1} \end{center}}  

\newcommand{\abstr}[1] {{\begin{center} \vskip .5cm {\bf Abstract
                        \vspace{0pt}} \end{center}}\begin{quote} #1
                        \end{quote}}

\newcommand{\e}{\varepsilon}

\newcommand{\trg}[3]{
\begin{picture}(35,30)(5,13)
\Line(5,5)(20,30)
\Line(35,5)(20,30)
\Line(5,5)(35,5)
\Line(5,5)(3,3)
\Line(35,5)(38,3)
\Line(20,30)(20,32)
\Text(20,0)[]{$\scriptstyle #1$}
\Text(13,25)[r]{$\scriptstyle #2$}
\Text(27,25)[l]{$\scriptstyle #3$}
\end{picture}
                    }

\newcommand{\trgleft}[3]{
\begin{picture}(35,30)(5,13)
\Line(5,5)(20,30)
\Line(35,5)(20,30)
\Line(5,5)(35,5)
\DashLine(5,5)(-10,5){2}
\Line(-10,5)(-13,3)
\Line(35,5)(38,3)
\Line(20,30)(20,32)
\Text(-3,0)[]{$\scriptstyle -1$}
\Text(20,0)[]{$\scriptstyle #1$}
\Text(13,25)[r]{$\scriptstyle #2$}
\Text(27,25)[l]{$\scriptstyle #3$}
\end{picture}
                        }

\begin{document}

\begin{titlepage}
\prepr{BI-TP-97/26\\ July 1997\\ }
\titul{The Differential equation method: \\
calculation of vertex-type diagrams with one non-zero mass }

\autor{J. Fleischer $^a$, A.V. Kotikov $^b$ and O.L. Veretin $^a$}
\lugar{$^a$ Fakult$\ddot{a}$t f$\ddot{u}$r Physik\\
Universit$\ddot{a}$t Bielefeld\\
D-33615 Bielefeld, Germany}
\lugar{
$^b$ Particle Physics Laboratory\\
Joint Institute for Nuclear Research\\
141980, Dubna (Moscow region), Russia}
\abstr{The differential equation method is applied to evaluate 
analytically two-loop vertex Feynman diagrams. Three on-shell 
infrared divergent planar two-loop diagrams with zero thresholds
contributing to the processes $Z \to b \overline b$ (for zero b mass)
and/or $H \to gg$ are calculated in order to demonstrate a new application
of this method.

\vskip 0.5 cm

}
\end{titlepage}
\newpage
\section{Introduction.}

In the two-loop order of the Standard Model large effects are due to the
exchange of heavy particles, like $H$ and $t$.
The present investigation is motivated by the need to perform accordingly
high precision
calculations of the decay rates $Z \to b \overline b$ and $H \to gg$.
We demonstrate a new application of 
the differential equation method (DEM) to calculate analytically several 
scalar vertex diagrams for the above decays. 
Intense study of the relevant diagrams is in 
progress (see  \cite{FT, FST, Fetal}) : semianalytically, using the technique
of asymptotic expansion of Feynman diagrams \cite{AS1, AS2} and subsequent
use of conformal mapping and summation by Pad\'{e} approximation.
We would like to note, however, that the technique of \cite{FT, FST} works 
best 
for diagrams with high thresholds. In the case of several massless virtual particles 
and in particular if the diagrams under consideration have zero thresholds,
additional contributions (apart from the ``naive'' diagram, see e.g. 
\cite{Fetal}) 
are needed due to the underlying method of the large mass expansion.
Moreover, in the above approach the convergence of the 
series w.r.t. the inverse masses is getting worse for large external momenta
squared the fewer heavy masses are involved. Thus, to obtain analytic results 
is quite an important task in particular for those cases with many massless 
virtual particles. If only one non-zero mass is involved, in fact the DEM
provides in a systematic manner these analytic results in a managable
amount of work. These analytic results in general are given in terms
of a one-dimensional integral representation from which it is possible to
obtain the Taylor series expansion in the external momentum squared. The
latter is considered as the most efficient approach for the final numerical
evaluation due to references \cite{FT, FST, Fetal}.

In the next section we shall recall the basic formulae and the sence of the 
DEM \cite{DEM1}-\cite{DEM3} (and \cite{DEMREW} 
for short review), which is used for the calculations. In Sections 3 and 4 
we shall demonstrate the basic steps in the
calculation of the massless two-loop diagram ($Case ~10$, according to the
notation of \cite{FST} and Fig. 1)
and one diagram with one non-zero mass ($Case ~7$). 
Section 5 contains only the results for another diagram ($Case ~8$).

\section{ The Differential equation method }

The DEM is a method which allows one to obtain results for massive
diagrams, evaluating only diagrams having an essentially simpler structure 
than the initial ones. 
The method is based on the rule of integration by part \cite{IBP}. 
Throughout the article we use the following notation. Assuming
dimensional regularization, all the calculations are performed in 
Euclidean momentum space of dimension $D=4-2\e$. The dotted and solid lines of
any  
diagram correspond to massless and massive euclidean propagators 
\footnote{The transition to Minkowsky space for a diagram $I \to \tilde I$
is given by changing $q_0 \to i\tilde q_0, ~q_j \to \tilde q_j$ $(j=1,2,3)$ 
($q$ and $ \tilde q$
are the external momenta of the diagram in Euclidean and Minkowsky space,
respectively).
Hereafter all variables with 
" $\tilde{}$ " 
denote Minkowsky space.
For the diagrams
under consideration, we will make these transitions at the final step 
of the calculation.}
 
\vspace{-5pt}\hfill\\
\hspace*{3cm}
$\frac{\displaystyle 1}{\displaystyle (q^2)^\alpha}\,=\;\;$
\begin{picture}(30,10)(5,4)
\DashLine(0,5)(30,5){2}
\Vertex(0,5){1}
\Vertex(30,5){1}
\Text(15,7)[b]{$\scriptstyle\alpha$} 
\end{picture} 
$,\qquad \frac{\displaystyle 1}{\displaystyle (q^2+m^2)^\alpha}\,=\;\;$
\begin{picture}(30,10)(5,4)
\Line(0,5)(30,5)
\Vertex(0,5){1}
\Vertex(30,5){1}
\Text(15,7)[b]{$\scriptstyle\alpha$} 
\Text(15,3)[t]{$\scriptstyle m^2$} 
\end{picture} 
\vspace{-0pt}\hfill\\
$\alpha$ and $m$ are called the index and mass of this line.
Further (except when mentioned otherwise) all solid lines have the 
same mass $m$. Lines with index 1 and mass $m$ are not marked.
Let us now consider the rules for our calculation.\\

{\bf Rule 1.} Massive tadpoles $ T(\alpha_1,\alpha_2)$
are integrated due to the graphic identity (hereafter we use the
Euclidean integral measure $d^Dk/{\pi}^{D/2}$)

\vspace{-15pt}\hfill\\
\hspace*{3cm}
$T(\alpha_1,\alpha_2)\,\equiv\;\;$
\begin{picture}(50,30)(0,13)
\DashCArc(20,15)(15,90,270){3}
\CArc(20,15)(15,-90,90)
\Vertex(20,30){1}
\Vertex(20,0){1}
\Line(20,0)(17,-3)
\Line(20,0)(23,-3)
\Text(4,18)[r]{$\scriptstyle\alpha_1$} 
\Text(38,18)[l]{$\scriptstyle\alpha_2$} 
\end{picture} 
$\,=\, \frac{\displaystyle R(\alpha_1,\alpha_2)}
        {\displaystyle (m^2)^{\alpha_1+\alpha_2-D/2}}$
\vspace{-0pt}\hfill\\
where
$$
R(\alpha_1,\alpha_2)~=~ 
\frac{\Gamma(D/2-\alpha_1)\Gamma(\alpha_1 +\alpha_2 -D/2)}{\Gamma(\alpha_2)
\Gamma(D/2)}  
$$
and $\Gamma$ is the Euler $\Gamma$-function.\\

{\bf Rule 2.} For a triangle the following recurrence relation \cite{IBP},
\cite{DEM1}-\cite{DEMREW} is
valid (here the line with index $\alpha_i$ has mass $m_i$)

\vspace{-10pt} \hfill \\
\hspace*{1cm}
\trg{\alpha_1}{\alpha_2}{\alpha_3}
$(D-2\alpha_1-\alpha_2-\alpha_3) \,=\, -2 m_1^2\alpha_1\,\,\,$
\trg{\alpha_1+1}{\alpha_2}{\alpha_3}
\vspace{20pt} \hfill \\
\vspace{-0pt}
\vspace{-20pt}\hfill \\
\hspace*{1cm}
$+\alpha_2\Biggl(\;\;$
\trg{\alpha_1-1}{\alpha_2+1}{\alpha_3}
$-\;\;\;\;\;$
\trgleft{\alpha_1}{\alpha_2+1}{\alpha_3}
$\,-(m_1^2+m_2^2)\;\;\;\;$
\trg{\alpha_1}{\alpha_2+1}{\alpha_3}
$\Biggr) + (\alpha_2\leftrightarrow\alpha_3)$
\vspace{+5pt} \hfill \\

We stress the fact that the basic line of the triangle always plays a
special role (in the following we call it ``distinctive'').

This Rule, i.e. the rule of integration by part, is obtained by multiplying
the integrand of the original diagram by $D=dk^{\mu}/dk^{\mu}$ ($k$ is
the momentum along the distinctive line) and using
$\int d^Dk ~ {\rm div}() =0$ for the regularized Feynman 
integrals.\\

{\bf Rule 3.} Massless loops $ L(q^2;\alpha_1,\alpha_2)$ with the external 
momentum $q$ are integrated due to the graphic identity

\vspace{-20pt}\hfill\\
\hspace*{3cm}
$L(q^2;\alpha_1,\alpha_2)\,\equiv\;\;$
\begin{picture}(60,30)(0,13)
\DashCurve{(5,15)(30,25)(55,15)}{2}
\DashCurve{(5,15)(30,5)(55,15)}{2}
\Line(5,15)(-5,15)
\Line(55,15)(65,15)
\Text(30,27)[b]{$\scriptstyle\alpha_1$} 
\Text(0,12)[t]{$q$} 
\Text(30,3)[t]{$\scriptstyle\alpha_2$} 
\end{picture} 
$\;\;=\;\;$
\begin{picture}(70,30)(0,4)
\DashLine(5,5)(65,5){2}
\Vertex(5,5){1}
\Vertex(65,5){1}
\Line(5,5)(-5,5)
\Line(65,5)(75,5)
\Text(33,7)[b]{$\scriptstyle\alpha_1+\alpha_2-D/2$} 
\Text(-3,-2)[b]{$q$} 
\end{picture} 
$\;\;A(\alpha_1,\alpha_2)$

\vspace{-5pt}\hfill\\

where
$$
A(\alpha_1,\alpha_2)~=~ 
\frac{a(\alpha_1) a(\alpha_2)}{a(\alpha_1 +\alpha_2 -D/2)}, ~~~
a(\alpha)~=~ 
\frac{\Gamma(D/2-\alpha)}{\Gamma(\alpha)}
$$

Stricly speaking, Rule 3 is not independent of Rules 1 and 2 because 
it may be obtained
(at least for even values of $\alpha_1$ or $\alpha_2$) in the limit $m^2 \to 0$
from equation (1) below. Hovewer, it is very convenient to 
consider it as third additional rule.\\

Consider a few specific examples.\\

{\bf 1.} A simple loop with one massive line

\vspace{-20pt}\hfill\\
\hspace*{3cm}
\begin{picture}(60,30)(0,13)
\DashCurve{(5,15)(30,25)(55,15)}{2}
\Curve{(5,15)(30,5)(55,15)}
\Line(5,15)(-5,15)
\Line(55,15)(65,15)
\Text(30,3)[t]{$\scriptstyle\alpha$} 
\Text(0,12)[t]{$q$} 
\end{picture} 
$\;\,=\,I_1(q^2;\alpha)$
\vspace{+5pt}\hfill\\

Applying Rule 2 ($\alpha_1=1, \alpha_2=\alpha, \alpha_3=0$) with the massless 
line as distinctive one, we get
$$
(D-2-\alpha) I_1(q^2;\alpha) ~=~ \alpha \Biggl[ T(0,\alpha +1)
-(q^2 + m^2) I_1(q^2;\alpha +1) \Biggr]
$$
The last diagram $-\alpha I_1(q^2;\alpha +1)$
on the r.h.s. is the derivative with respect to $m^2$ 
of the initial diagram $I_1(q^2;\alpha )$. Hence, the r.h.s. has the form
$$
\alpha R(0,1+\alpha)\frac{1}{{\left[m^2 \right]}^{\alpha +1 -D/2}}
  +(q^2 + m^2) \frac{d}{dm^2}I_1(q^2;\alpha),
$$
Thus we get a differential equation with respect to $m^2$ for the initial 
 diagram. Its solution for $I_1$ is
$$
q^{2(\alpha+1-D/2)} I_1(q^2;\alpha) ~=~ \frac{1}{t^{D-2-\alpha}} \int^t_0 ds
\frac{\alpha R(0,1+\alpha)}{s^{2-D/2}(1-s)^{\alpha+1-D/2}},~~~~~
(t=
\frac{q^2}{q^2 + m^2})
$$
Notice
that sometimes a more convenient representation of the diagram $I_1$
is given in the form

\vspace{-20pt}\hfill\\
\hspace*{1cm}
\begin{picture}(60,30)(0,13)
\DashCurve{(5,15)(30,25)(55,15)}{2}
\Curve{(5,15)(30,5)(55,15)}
\Line(5,15)(-5,15)
\Line(55,15)(65,15)
\Text(30,3)[t]{$\scriptstyle\alpha$} 
\Text(0,12)[t]{$q$} 
\end{picture} 
$\displaystyle   \;\,=\,\alpha\, R(0,\alpha+1) 
    \int_0^1 \frac{ds}{(1-s)^{2-D/2}s^{\alpha+1-D/2}}\;\;\;$
\begin{picture}(70,30)(0,4)
\Line(5,5)(65,5)
\Vertex(5,5){1}
\Vertex(65,5){1}
\Line(5,5)(-5,5)
\Line(65,5)(75,5)
\Text(33,7)[b]{$\scriptstyle\alpha+1-D/2$} 
\Text(33,3)[t]{$\scriptstyle m^2/s$} 
\Text(-3,-2)[b]{$q$} 
\end{picture}
\hfill (1)\\ 
\vspace{+15pt}

{\bf 2.} A simple loop with two massive lines

\vspace{-20pt}\hfill\\
\hspace*{3cm}
\begin{picture}(60,30)(0,13)
\Curve{(5,15)(30,25)(55,15)}
\Curve{(5,15)(30,5)(55,15)}
\Line(5,15)(-5,15)
\Line(55,15)(65,15)
\Text(0,12)[t]{$q$} 
\end{picture} 
$\;\,=\,I_2(q^2)$
\vspace{+5pt}\hfill\\

In analogy with the previous consideration we get

\vspace{-15pt}\hfill\\
\hspace*{1cm}
\begin{picture}(60,30)(0,13)
\Curve{(5,15)(30,25)(55,15)}
\Curve{(5,15)(30,5)(55,15)}
\Line(5,15)(-5,15)
\Line(55,15)(65,15)
\Text(0,12)[t]{$q$} 
\end{picture} 
$\displaystyle   \;\,=\,2^{3-D}\, R(0,2) 
    \int_0^1 \frac{ds}{(1-s)^{1/2}s^{2-D/2}}\;\;\;$
\begin{picture}(70,30)(0,4)
\Line(5,5)(65,5)
\Vertex(5,5){1}
\Vertex(65,5){1}
\Line(5,5)(-5,5)
\Line(65,5)(75,5)
\Text(33,7)[b]{$\scriptstyle 2-D/2$} 
\Text(33,3)[t]{$\scriptstyle 4m^2/s$} 
\Text(-3,-2)[b]{$q$} 
\end{picture}
\hfill  (2)\\ 
\vspace{-0pt}

It is clearly seen that Eqs.(1) and (2) allow us to reduce 
an $l$-loop diagram to an
$(l-1)$-loop diagram with one propagator 
having "mass" $m^2/s$ and $4m^2/s$, respectively.
Applications of these relations may be found in
\cite{DEM1, DEM2}.

Notice that if some lines have index 1,
Rule 2 leads to a differential equation for
the initial diagram with simpler diagrams in the inhomogeneous term.
If, however,all indices are different from 1,
Rule 2 does not lead to a simplification of the inhomogeneous term.
In this case a more complicated technique, for example 
the method of Feynman parameters, is needed.
Using the method of Feynman parameters the
relation between $l$-loop and $(l-1)$-loop diagrams in general may be
obtained in the form (as before $\alpha_i$ corresponds to the mass $m_i$):

\vspace{-10pt}\hfill\\
\hspace*{0.1cm}
\begin{picture}(60,30)(0,13)
\Curve{(5,15)(30,25)(55,15)}
\Curve{(5,15)(30,5)(55,15)}
\Line(5,15)(-5,15)
\Line(55,15)(65,15)
\Text(30,27)[b]{$\scriptstyle\alpha_1,\,m_1$} 
\Text(30,3)[t]{$\scriptstyle\alpha_2,\,m_2$} 
\Text(0,12)[t]{$q$} 
\end{picture} 
$\displaystyle   \;\,=\,
   \frac{\Gamma(\alpha_1+\alpha_2-D/2)}{\Gamma(\alpha_1)\,\Gamma(\alpha_2)}
   \int_0^1 \frac{ds}{(1-s)^{\alpha_1+1-D/2}s^{\alpha_2+1-D/2}}\;\;\;$
\begin{picture}(70,30)(0,4)
\Line(5,5)(65,5)
\Vertex(5,5){1}
\Vertex(65,5){1}
\Line(5,5)(-5,5)
\Line(65,5)(75,5)
\Text(33,7)[b]{$\scriptstyle\alpha_1+\alpha_2-D/2$} 
\Text(33,3)[t]{$\scriptstyle \frac{m_1^2}{1-s}+\frac{m_2^2}{s}$} 
\Text(-3,-2)[b]{$q$} 
\end{picture}
\hfill (3)\\ 
\vspace{+10pt}\\
   For $\alpha_1=1$ and $\alpha_2=\alpha$ and $m_1=0$ (3) reduces to (1).\\

{\bf 3.} Vertex-type diagrams and $N$-point functions were studied in 
\cite{DEM2} and \cite{DEM3}, respectively. Further simple examples
are worked out in these Refs. Since the basic subject of our 
investigation is to
calculate vertex diagrams in $Z \to b \overline b$ and $H 
\to gg$ decays, we consider here as another example the one-loop
vertex diagram with the kinematics $q_1^2=q^2_2=0, ~(q_1-q_2)^2=q^2$: 

\vspace{-0pt}\hfill\\
\hspace*{3cm}
\begin{picture}(35,30)(5,13)
\DashLine(5,5)(20,30){2}
\DashLine(35,5)(20,30){2}
\Line(5,5)(35,5)
\Line(5,5)(0,0)
\Line(35,5)(40,0)
\Line(20,30)(20,35)
\Text(18,32)[r]{$q_1-q_2$}
\Text(0,3)[r]{$q_2$}
\Text(40,3)[l]{$q_1$}
\end{picture}
$\;\;=\;I_3(q^2,m^2)$
\vspace{10pt}\hfill\\

Applying Rule 2 with the massive and one of the  massless as distictive 
lines, we obtain, respectively

\vspace{-20pt}\hfill\\
\hspace*{3cm}
$(D-4)I_3\,=\,  2L(q^2,1,2) - 2m^2\!\!$
\begin{picture}(35,30)(5,13)
\DashLine(5,5)(20,30){2}
\DashLine(35,5)(20,30){2}
\Line(5,5)(35,5)
\Line(5,5)(3,3)
\Line(35,5)(38,3)
\Line(20,30)(20,32)
\Text(27,25)[l]{$\scriptstyle 2$}
\end{picture}
$ - 2m^2\!\!$
\begin{picture}(35,30)(5,13)
\DashLine(5,5)(20,30){2}
\DashLine(35,5)(20,30){2}
\Line(5,5)(35,5)
\Line(5,5)(3,3)
\Line(35,5)(38,3)
\Line(20,30)(20,32)
\Text(20,0)[]{$\scriptstyle 2$}
\end{picture}
\hfill (4)\\
\vspace{10pt}
\vspace{-10pt}\hfill\\
\hspace*{3cm}
$(D-4)I_3\,=\,  T(1,2) + T(2,1)- q^2\!\!$
\begin{picture}(35,30)(5,13)
\DashLine(5,5)(20,30){2}
\DashLine(35,5)(20,30){2}
\Line(5,5)(35,5)
\Line(5,5)(3,3)
\Line(35,5)(38,3)
\Line(20,30)(20,32)
\Text(27,25)[l]{$\scriptstyle 2$}
\end{picture}
$ - m^2\!\!$
\begin{picture}(35,30)(5,13)
\DashLine(5,5)(20,30){2}
\DashLine(35,5)(20,30){2}
\Line(5,5)(35,5)
\Line(5,5)(3,3)
\Line(35,5)(38,3)
\Line(20,30)(20,32)
\Text(20,0)[]{$\scriptstyle 2$}
\end{picture}
\hfill (5)\\
\vspace{10pt}

{\bf a)} For the massless case, Eq.(4) leads to the result for $I_3(q^2, 0)$
in the form
$$
I_3(q^2, 0)~=~ -\frac{1}{\e} A(1,2) \frac{1}{q^{2(1+\e)}} ~=~
\frac{1}{\e^2} 
\frac{\Gamma(1+\e)}{q^{2(1+\e)}} \frac{\Gamma^2(1 - \e)}{\Gamma(1-2\e)}
$$\\

{\bf b)} For the massive case combining  (4) and (5), we get
the differential equation for $I_3(q^2, m^2)$ 
in the form
\setcounter{equation}{5}
\begin{eqnarray}
\biggl(1-\frac{2m^2}{q^2}\biggr)(D-4) I_3  ~=~  2f_3 +
2m^2 \biggl(1-\frac{m^2}{q^2}\biggr) \frac{d}{dm^2}I_3,  \nonumber
\end{eqnarray}
where
$$
f_3 ~=~ L(q^2,1,2) - \frac{m^2}{q^2} \biggl( T(1,2) + T(2,1) \biggr)
$$
is a combination of the massless loop and massive tadpoles.

Introducing the variable $x=q^2/m^2$ and using the boundary condition:
$I_3=0$ for $m^2 \to \infty$, we get
\begin{eqnarray}
I_3(q^2, m^2)&=& \frac{1}{\e} \frac{\Gamma(1+\e)}{q^{2(1+\e)}}
\frac{x^{2\e}}{(1-x)^{\e}} \int^x_0 \frac{dy~ y^{-2\e}}{(1-y)^{1-\e}}
\Biggl[ \frac{\Gamma^2(1 - \e)}{\Gamma(1-2\e)} - y^{\e} \Biggr] 
\nonumber \\
&=&
\frac{1}{q^2} \biggl[\zeta(2) - {\rm Li}_2(1-x)\biggr],
\label{6} 
\end{eqnarray}
where
${\rm Li}_{n}$ is the polylogarithm 
and $\zeta(n)$ the Riemann  $\zeta$ - function.

In Minkowsky space (\ref{6}) reads

\begin{eqnarray}
\tilde I_3(\tilde{q}^2, m^2)~=~ \frac{i}{\tilde{q}^2} 
\biggl[\zeta(2) - {\rm Li}_2(1+\tilde{x})\biggr],
\nonumber 
\end{eqnarray}
where $\tilde{x} = \tilde{q}^2/m^2$. The appearence of an imaginary 
part for $i I_3$ from ${\rm Li}_2(1+ \tilde{x})$ for $\tilde{x}>0$ is
seen from the analytic properties of ${\rm Li}_2$ .\\

The key idea of the DEM: using the rule of integration by part (Rule 2)
with different distinctive lines, the results for
massive Feynman diagrams are obtained as a differential equation with an
inhomogeneous term\footnote{Sometimes (see Eq.(8) below, for example)
 the differential equation can be
brougt to a simple arithmetic
relation between the initial diagram and simpler diagrams
in the inhomogeneous term.} containing simpler diagrams.
Using this procedure repeatedly one can reduce the 2-loop 
diagrams to 1-loops. These are integrated either 
again with the help of Rule 2 (by analogy with example 3),
decreasing step by step the number of lines, or using the 
method of Feynman parameters for more complicated cases.

The main difference from the massless case is the necessity to integrate the 
final result with respect to the mass several times. 
The basic advantage compared to the usual methods (for example, Feynman 
parameter method) to calculate massive Feynman diagrams is the appearence
of complicated functions in the final step only. Hence, these functions do 
not interfere the process of calculation.

\section{ Two-loop vertex diagrams}

At two-loop level we consider here three diagrams (see Fig. 1) out of ten
characterized in \cite{FST}. They have the kinematics: 
$q_1^2=q^2_2=0, ~(q_1-q_2)^2=q^2$.
The evaluation of the 
massless diagram $J_{10}$
 ($Case ~10$) and one diagram with a non-zero mass $J_7$ ($Case ~7$) will be 
fully presented. For another diagram ($J_8$, i.e. 
$Case ~8$) we qive only the final result. 

Applying Rule 2 with line 6 (see Fig. 1) as distinctive one, we
have for the $J_7$ diagram:

\vspace{-30pt}\hfill\\
\hspace*{3cm}
$(D-4)J_7\,=\, 2\!\!$
\begin{picture}(35,60)(5,27)
\DashLine(5,55)(35,55){2}
\DashLine(20,30)(5,55){2}
\DashLine(20,30)(35,55){2}
\BCirc(20,17.5){12.5}
\Line(20,5)(20,3)
\Line(5,55)(3,58)
\Line(35,55)(38,58)
\Text(35,15)[l]{$\scriptstyle 2$}
\end{picture}
$\!\! -\,2\;$
\begin{picture}(35,60)(5,27)
\DashLine(5,55)(35,55){2}
\DashLine(5,30)(5,55){2}
\Line(5,30)(35,55)
\Line(20,5)(5,30)
\Line(20,5)(35,55)
\Line(20,5)(20,3)
\Line(5,55)(3,58)
\Line(35,55)(38,58)
\Text(25,15)[l]{$\scriptstyle 2$}
\end{picture}
$ -\,4m^2$
\begin{picture}(35,60)(5,27)
\DashLine(5,55)(35,55){2}
\DashLine(5,30)(5,55){2}
\DashLine(35,30)(35,55){2}
\Line(5,30)(35,30)
\Line(20,5)(5,30)
\Line(20,5)(35,30)
\Line(20,5)(20,3)
\Line(5,55)(3,58)
\Line(35,55)(38,58)
\Text(29,15)[l]{$\scriptstyle 2$}
\end{picture}
$ -\,2m^2$
\begin{picture}(35,60)(5,27)
\DashLine(5,55)(35,55){2}
\DashLine(5,30)(5,55){2}
\DashLine(35,30)(35,55){2}
\Line(5,30)(35,30)
\Line(20,5)(5,30)
\Line(20,5)(35,30)
\Line(20,5)(20,3)
\Line(5,55)(3,58)
\Line(35,55)(38,58)
\Text(20,34)[l]{$\scriptstyle 2$}
\end{picture}
\hfill (7)\\ 
\vspace{15pt}

Repeating the application of Rule 2 to the second 
diagram of Eq.(7) with the left vertical line as distinctive one,
we get

\vspace{-30pt}\hfill\\
\hspace*{3cm}
$(D-4)$
\begin{picture}(35,60)(5,27)
\DashLine(5,55)(35,55){2}
\DashLine(5,30)(5,55){2}
\Line(5,30)(35,55)
\Line(20,5)(5,30)
\Line(20,5)(35,55)
\Line(20,5)(20,3)
\Line(5,55)(3,58)
\Line(35,55)(38,58)
\Text(25,15)[l]{$\scriptstyle 2$}
\end{picture}
$=$
\begin{picture}(35,60)(5,23)
\Line(20,5)(5,40)
\Line(20,5)(35,40)
\Curve{(5,40)(20,33)(35,40)}
\DashCurve{(5,40)(20,47)(35,40)}{2}
\Line(20,5)(20,3)
\Line(5,40)(3,43)
\Line(35,40)(38,43)
\Text(28,15)[l]{$\scriptstyle 2$}
\Text(20,49)[b]{$\scriptstyle 2$}
\end{picture}
$+$
\begin{picture}(35,60)(5,23)
\Line(20,5)(5,40)
\Line(20,5)(35,40)
\Curve{(5,40)(20,33)(35,40)}
\DashCurve{(5,40)(20,47)(35,40)}{2}
\Line(20,5)(20,3)
\Line(5,40)(3,43)
\Line(35,40)(38,43)
\Text(28,15)[l]{$\scriptstyle 2$}
\Text(20,31)[t]{$\scriptstyle 2$}
\end{picture}
$-$
\begin{picture}(35,60)(5,23)
\DashLine(20,5)(5,40){2}
\DashLine(5,40)(35,40){2}
\Curve{(20,5)(25,27)(30,36)(35,40)}
\Curve{(20,5)(25,9)(30,17)(35,40)}
\Line(20,5)(20,3)
\Line(5,40)(3,43)
\Line(35,40)(38,43)
\Text(22,25)[r]{$\scriptstyle 2$}
\Text(32,15)[l]{$\scriptstyle 2$}
\end{picture}
\hfill (8)\\
\vspace{25pt}

{\bf  Case 10.} 
For the massless case, Eqs.(7) and (8) lead to the result
for $J_{10}$ in the following form 

\vspace{-15pt}\hfill\\
\hspace*{1cm}
$(-2\e)J_{10}\,= \,2A(1,2) 
\Biggl( \frac{\displaystyle 1}{\displaystyle q^{2(1+\epsilon)}}$
\begin{picture}(35,30)(5,13)
\DashLine(20,5)(5,30){2}
\DashLine(20,5)(35,30){2}
\DashLine(5,30)(35,30){2}
\Line(20,5)(20,3)
\Line(5,30)(3,33)
\Line(35,30)(38,33)
\end{picture}
$+\,\frac{\displaystyle 1}{\displaystyle \epsilon}$
\begin{picture}(35,30)(5,13)
\DashLine(20,5)(5,30){2}
\DashLine(20,5)(35,30){2}
\DashLine(5,30)(35,30){2}
\Line(20,5)(20,3)
\Line(5,30)(3,33)
\Line(35,30)(38,33)
\Text(20,33)[b]{$\scriptstyle 1+\epsilon$}
\Text(27,13)[l]{$\scriptstyle 2$}
\end{picture}
$\Biggr)-\frac{\displaystyle 1}{\displaystyle \epsilon}A(2,2)$
\begin{picture}(35,30)(5,13)
\DashLine(20,5)(5,30){2}
\DashLine(20,5)(35,30){2}
\DashLine(5,30)(35,30){2}
\Line(20,5)(20,3)
\Line(5,30)(3,33)
\Line(35,30)(38,33)
\Text(27,13)[l]{$\scriptstyle 2+\epsilon$}
\end{picture}
\hfill (9)\\
\vspace{+5pt}

The first diagram on the r.h.s. of Eq.(9) was already calculated in the
previous section, the third one may be done in an analog manner. The second 
diagram 
may be evaluated by the method of Feynman parameters. The result for $J_{10}$
is known (see \cite{Neervan}) and has the form:
\addtocounter{equation}{3}
\begin{eqnarray}
J_{10}&=& \frac{1}{\e^4} \frac{\Gamma(1+\e)}{{(q^2)}^{2(1+\e)}}
\frac{\Gamma^2(1 - \e)}{\Gamma(1-2\e)} \Biggl[
\frac{\Gamma^2(1 - \e)\Gamma(1+\e)}{\Gamma(1-2\e)} ~-~ 
\frac{3}{2} \frac{\Gamma(1-2\e)\Gamma(1+2\e)\Gamma(1-\e)}{\Gamma(1-3\e)
\Gamma(1+\e)} 
 \nonumber \\ &+&
\frac{3}{4} \frac{\Gamma^2(1-2\e)\Gamma(1+2\e)}{\Gamma(1-3\e)} \Biggr]
 \label{a} \\
&=& \frac{1}{4\e^4} \frac{\Gamma^2(1+\e)}{q^{2(1+\e)}}
\Biggl[ 1+ 4\zeta(2)\e^2 + 20\zeta(3)\e^3 + 22\zeta(4)\e^4   \Biggr]
~+~ O(\e) \nonumber 
\end{eqnarray}
We consider our derivation particularly simple.
In Minkowsky space the diagram $\tilde J_{10}$ is represented in the form
(10) with the replacement $q^2 \to - \tilde{q}^2$ and 
the additional factor $-1$. \\

{\bf Case 7.} 
For the case with one non-zero mass, Eq.(7) leads to a differential 
equation for $J_{7}$ since the sum of the last two diagrams 
on the r.h.s. is the derivative of the initial diagram $J_7$ with respect 
to $m^2$. Solving it, we have
\begin{eqnarray}
 J_7 ~=~ x^{\e} \int^x_0 \frac{dy}{y^{1+\e}}f_7,    \label{14} 
\end{eqnarray}
where $f_7$ is the inhomogeneous term containing the two first diagrams 
of the r.h.s. of Eq.(7). These diagrams are
effectively only one-loop vertex diagrams. 
Indeed, using for the second one Eq.(3) with $m_1^2 =0$ to the first two 
diagrams of
the r.h.s. of Eq.(8) and with $m_1^2 = m_2^2$ to the last one,
we have 

\vspace{-25pt}\hfill\\
\hspace*{0.5cm}
\begin{picture}(35,60)(5,27)
\DashLine(5,55)(35,55){2}
\DashLine(5,30)(5,55){2}
\Line(5,30)(35,55)
\Line(20,5)(5,30)
\Line(20,5)(35,55)
\Line(20,5)(20,3)
\Line(5,55)(3,58)
\Line(35,55)(38,58)
\Text(25,15)[l]{$\scriptstyle 2$}
\end{picture}
$\displaystyle \,=\,\, -\frac{\Gamma(1+\epsilon)}{2\epsilon}
  \int_0^1\frac{ds}{\bigr( s(1-s) \bigl)^{1+\epsilon}}\; \Biggl(\;\;\;$
\begin{picture}(35,30)(5,18)
\Line(20,5)(0,40)
\Line(20,5)(40,40)
\Line(0,40)(40,40)
\Line(20,5)(20,3)
\Line(0,40)(-3,43)
\Line(40,40)(43,43)
\Text(20 ,43)[b]{$\scriptstyle 1+\epsilon$}
\Text(20,38)[t]{$\scriptstyle m^2\!/s$}
\Text(37,24)[l]{$\scriptstyle 2$}
\end{picture}
$-\,(1+\epsilon)$
\begin{picture}(35,30)(5,18)
\DashLine(20,5)(0,40){2}
\Line(20,5)(40,40)
\DashLine(0,40)(40,40){2}
\Line(20,5)(20,3)
\Line(0,40)(-3,43)
\Line(40,40)(43,43)
\Text(37,24)[l]{$\scriptstyle 2+\epsilon$}
\Text(27,12)[l]{$\scriptstyle m^2\!/s(1-s)$}
\end{picture}
~~~~~~~~~$\Biggr)$
\hfill (12)\\
\vspace{20pt}

Evaluating the one-loop diagrams on the r.h.s. of Eq.(12) by
the method of Feynman parameters and
calculating the one-loop parts of the first diagram on the r.h.s. of 
Eq.(7), we obtain the final result for 
the diagram $J_7$ from Eq.(\ref{14})
(neglecting terms $\sim O(\e)$)
\addtocounter{equation}{1}
\begin{eqnarray}
J_{7}&=& \frac{\Gamma^2(1+\e)}{2{(q^2)}^2 {(m^2)}^{2\e}}
\int^1_0 \frac{ds}{s(1-s)} 
\Biggl[ \frac{1}{\e^2} \ln(1+z) 
+ \frac{1}{\e} \biggl(\frac{1}{2}\ln^2(1+z) - \ln(x)\ln(1+z) \biggr) 
\nonumber \\
&+& \Phi^{(7)}_0(z)+ \frac{1}{2} \ln(x)\ln^2(1+z) + \frac{1}{2}\ln^2(x)\ln(1+z)
  \Biggr],
 \label{12} 
\end{eqnarray}
where $z=xs(1-s)$, 
\begin{eqnarray}
 \Phi^{(7)}_0(z) &=& \ln[s(1-s)]\ln^2(1+z) - \frac{7}{6} \ln^3(1+z)
 - 12 S_{1,2}(-z) - 5 \ln(1+z){\rm Li}_2(-z) 
 \nonumber \\
&-& \zeta(2)\ln(1+z)
\nonumber
\end{eqnarray}
and hereafter \cite{DD}
\begin{eqnarray}
S_{n,k}(x) ~=~  \frac{(-1)^{n+k-1}}{(n-1)!k!} \int^1_0 
\frac{dy}{y} \ln^{n-1}(y) \ln^k(1-xy),
~~ 
S_{n,1}(x) ~=~ {\rm Li}_{n+1}(x)
 \nonumber
\end{eqnarray}\\

{\bf 1.} In Refs. \cite{FT, FST, Fetal} some diagrams 
contributing to the processes $Z \to b \overline b$
and $H \to gg$ were evaluated semianalytically, i.e. the
first coefficients of the Taylor series in $q^2/m^2$  were calculated
analytically (in terms of rational numbers and $\zeta(2)$, $\zeta(3)$) 
and the diagram itself is 
reconstructed by means of conformal mapping and the
Pad\'{e} approximation technique. 
It turned out that this
approach is particularly useful for the final numerical
evaluation since it allows in general to calculate the diagram to higher 
precision than the numerical integration. Therefore we give below
the Taylor series of (13). The importance of the 
obtained formula below is that arbitrary many Taylor coefficients can
be obtained very easily while in Ref.\cite{Fetal} their calculation
needed much computertime. This means that with such a representation 
the diagram can be
calculated numerically to even higher precision with almost no effort.

\begin{eqnarray}
J_{7}&=& -\frac{\Gamma^2(1+\e)}{{(q^2)}^2 {(m^2)}^{2\e}}
\sum^{\infty}_{n=1}  
\frac{(-x)^n \Gamma^2(n)}{\Gamma(2n+1)}
\Biggl[\frac{1}{\e^2} 
~-~ \frac{1}{\e} \biggl(\ln(x) +S_1(n-1)  \biggr) - \frac{3}{2}S_2(n-1)  
\nonumber \\ &-&   \frac{15}{2}S^2_1(n-1) 
 +4S_1(n-1)S_1(2n) - \zeta(2)
 - \ln(x)S_1(n-1) + \frac{1}{2}\ln^2(x)
  \Biggr],
 \nonumber
\end{eqnarray}\\
where
$$
 S_l(n) ~=~  \sum^{n}_{1}\frac{1}{k^l}
$$\\ 

{\bf 2.} In Minkowsky space the diagram $\tilde J_7$ is represented in 
the form (13) with the replacement $x \to - \tilde{x}$ and the 
additional factor $-1$.
The imaginary part generates from the terms 
$\ln(-\tilde{x})=\ln(\tilde{x}) - i\pi$ and 
$\ln^2(-\tilde{x})$ below the threshold at $\tilde{q}^2=4m^2$ and 
from the  functions
$S_{n,k}(\tilde{x}s(1-s))$ above. 

At threshold $\tilde{q}^2=4m^2$
we can get a closed form for diagram $\tilde J_7$. 
Splitting $\tilde J_7$ into real and 
imaginary part: $ \tilde J_{7}~=~ {\rm Re}\,\tilde J_{7} +i {\rm Im}\,
\tilde J_{7} $,  
we have
\begin{eqnarray}
{\rm Re}\,\tilde J_{7}&=&  \frac{\Gamma^2(1+\e)}{16{(m^2)}^{2+2\e}}  \Biggl[
\frac{3}{\e^2} \zeta(2) ~-~ \frac{1}{\e} \biggl(7\zeta(3) +6 \zeta(2)\ln(2)
\biggr)
\nonumber \\ 
&+& 6\zeta(2)\ln^2(2) - 14\zeta(3)\ln(2) 
- \frac{511}{4}\zeta(4) + 16 U_{3,1}  \Biggr] \nonumber \\
\frac{1}{\pi} {\rm Im}\,\tilde J_{7}&=&  
\frac{\Gamma^2(1+\e)}{16{(m^2)}^{2+2\e}}
\Biggl[ \frac{3}{\e} \zeta(2)  - 6 \zeta(2)\ln(2)
+ 7\zeta(3)  \Biggr],
 \nonumber 
\end{eqnarray}
where the function
$$
U_{3,1} ~=~  \sum_{n>m>0}\frac{(-1)^{n+m}}{n^3m}~=~
\frac{1}{2}\zeta(2)\ln^2(2) - \frac{1}{12}\ln^4(2)
+ \frac{1}{2}\zeta(4) - 2 {\rm Li}_{4}(\frac{1}{2})
$$
was defined by Broadhurst in \cite{BRO}.\\

{\bf Case 8.} The scheme for calculating diagram $J_8$ is
similar to the one of the previous subsection. 
Here we will present 
only the final result for this diagram.\\
 
{\bf 1.}The result of the diagram $J_8$ may be given in the following
form 
\begin{eqnarray}
J_{8}&=&  \frac{\Gamma^2(1+\e)}{{(q^2)}^2 {(m^2)}^{2\e}}
\Biggl[ \frac{1}{\e^2} \Phi^{(8)}_2(x) 
+ \frac{1}{\e} \Phi^{(8)}_1(x)
+ \Phi^{(8)}_0(x)
  \Biggr],
 \label{24} 
\end{eqnarray}\\
where
$\Phi^{(8)}_1(x) = \overline \Phi^{(8)}_1(x) + \tilde \Phi^{(8)}_1(x)$
\mbox{ and }
$\Phi^{(8)}_0(x) = \overline \Phi^{(8)}_0(x) + \tilde \Phi^{(8)}_0(x)$\\
and
\begin{eqnarray}
\Phi^{(8)}_2(x) &=& \zeta(2) - {\rm Li}_2(1-x), \nonumber \\
\overline
\Phi^{(8)}_1(x) &=& 3 {\rm Li}_3(1-x) - 3 S_{1,2}(1-x) - 
2 \zeta(2) \ln(1-x), \nonumber \\
\tilde \Phi^{(8)}_1(x) &=& 4 \int^x_0 \frac{dy}{1-y}{\rm Li}_2(-y) 
\nonumber \\&=&
\biggl[ \ln(1-x){\rm Li}_2(-x) + \frac{1}{2} S_{1,2}(x^2) - S_{1,2}(x) -
S_{1,2}(-x) \biggr],   \nonumber \\
\overline \Phi^{(8)}_0(x) &=& 2 \zeta(3) -
7 S_{1,3}(1-x) -9 {\rm Li}_4(1-x) +11 S_{2,2}(1-x) -\frac{3}{2} 
{\rm Li}^2_2(1-x) 
\nonumber \\
&+&  \zeta(2)  \biggl[ \ln^2(1-x) +6 {\rm Li}_2(1-x)  \biggl]  
 \nonumber \\
\tilde \Phi^{(8)}_0(x) &=& 4 \int^x_0 \frac{dy}{1-y} \biggl[
\frac{3}{2} {\rm Li}_3(-y) +4S_{1,2}(-y) +\ln(1-y){\rm Li}_2(-y) \biggr]
\nonumber \\
&-& \tilde \Phi^{(8)}_1(x)  \ln(1-x)
\nonumber 
\end{eqnarray}
or the expansion w.r.t. $q^2/m^2$:
\begin{eqnarray}
J_{8}&=& \frac{\Gamma^2(1+\e)}{{(q^2)}^2 {(m^2)}^{2\e}}
\sum^{\infty}_{n=1} \frac{x^n}{n}
\Biggl[\frac{1}{\e^2} \biggl(\frac{1}{n} - \ln(x) \biggr)
~+~ \frac{1}{\e} \biggl(\tilde \phi^{(8)}_1(n) - 3\ln(x) \Bigl[
S_1 + \frac{1}{n} \Bigr] + \frac{3}{2}\ln^2(x) \biggr) 
\nonumber \\
&+& \phi^{(8)}_0(n)- \ln(x)\phi^{(8)}_1(n) + \frac{1}{2} \ln^2(x) \Bigl[
5S_1 + \frac{7}{n} \Bigr] - \frac{7}{6}\ln^3(x)
  \Biggr],
 \nonumber 
\end{eqnarray}\\
where
\begin{eqnarray}
\tilde \phi^{(8)}_1(n) &=& 3S_2 -4K_2 +\frac{3S_1}{n}+ \frac{3}{n^2}
- \zeta(2), \nonumber \\
\phi^{(8)}_1(n) &=& \frac{7}{2}S_2 + \frac{9}{2}S^2_1 
 +\frac{5S_1}{n}+ \frac{7}{n^2}
- 3\zeta(2)
 \nonumber \\
 \phi^{(8)}_0(n) &=&  9S_1S_2 + 2S_3 - 2K_3 - 12K_{2,1}  - 4S_1K_2
+ \frac{9S_1^2 + 7S_2}{2n} + 5\frac{S_1}{n^2} + \frac{7}{n^3} \nonumber \\
&-&  \zeta(2)\Bigl[ 7S_1 + \frac{3}{n} \Bigr] - 2 \zeta(3) 
\nonumber 
\end{eqnarray}
and

$$
S_l \equiv S_l(n-1),~~ K_l~=~  \sum^{n-1}_{1}\frac{(-1)^{k+1}}{k^l}
~ \mbox{ and }~
K_{2,1} 
~=~  \sum^{n-1}_{1}\frac{(-1)^{k+1}}{k^2}S_1(k-1)
$$\\

{\bf 2.} In Minkowsky space the diagram $\tilde J_8$
is represented in the form
(\ref{24}) with the replacement $x \to - \tilde{x}$ and the additional 
factor $-1$.
The imaginary part generates from the  functions
$S_{n,k}(1+\tilde{x})$, when 
$\tilde{x}>0$.

The calculation of the diagram $\tilde J_8 ={\rm Re}\,\tilde J_{8} 
+i {\rm Im}\,\tilde J_{8}$ 
at threshold $\tilde{q}^2=m^2$ leads to
the following result
\begin{eqnarray}
{\rm Re}\,\tilde J_{8}&=&  \frac{\Gamma^2(1+\e)}{{(m^2)}^{2+2\e}}  \Biggl[
\frac{1}{2\e^2} \zeta(2) \nonumber \\
&+& \frac{1}{2\e} \biggl( \frac{1}{4}\zeta(3) -15 \zeta(2)\ln(2)
\biggr) 
+ \frac{45}{4}\zeta(2)\ln^2(2) 
- \frac{401}{16}\zeta(4) + \frac{15}{2} U_{3,1}  \Biggr] \nonumber \\
\frac{1}{\pi} {\rm Im}\,\tilde J_{8}&=&  \frac{\Gamma^2(1+\e)}{{(m^2)}^{2+2\e}}
\Biggl[\frac{1}{\e^2} \ln(2) \nonumber \\
&+& \frac{3}{2\e} \biggl( \zeta(2) - \ln^2(2)  \biggr) 
+ \frac{3}{2}\ln^3(2) - 14 \zeta(2)\ln(2)
+ \frac{53}{8}\zeta(3)  \Biggr]
 \nonumber
\end{eqnarray}\\

\section{ Conclusion}

The differential equation method has been applied to 
demonstrate a new approach to calculate two-loop
vertex-type diagrams with one non-zero mass. Analytic
results have been obtained for several scalar two-loop diagrams
of the processes $Z \to b \overline b$ 
and $H \to gg$, which will be relevant for tests 
of the Standard Model. As main result we obtained a 
closed formula for the Taylor coefficients
for the diagrams under consideration, which allows their easy numerical
evaluation with extremely high precision. The obtained coefficients
were checked against those of Ref. \cite{Fetal}, which were, however,
only known explicitely up to a certain highest index in terms of rational 
numbers and $\zeta(2)$,$\zeta(3)$.

\vskip 0.5 cm
%
%
%
%

This work was supported in part by the Volkswagen Stiftung under I/71~293
and the BMBF under 05~7BI92P~9.
We are grateful to M. Kalmykov, V. Smirnov and M. Tentyukov for discussions.

\vskip 0.5 cm

%
%

%
\newpage
%
%

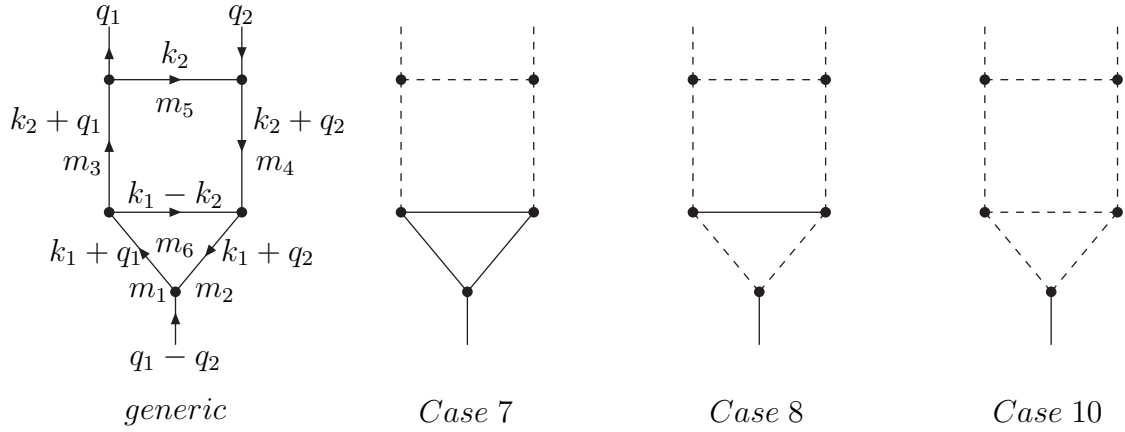
{\bf Figure 1.} Planar diagrams with zero threshold occuring
in $Z \to b \overline b$ and $H \to gg$ according to notation of
Ref. \cite{Fetal} . Dashed lines are massless and solid 
lines massive.


\vspace{1cm}
\vspace*{1.0cm}
\begin {figure} [htbp]
\begin{picture}(400,160)(-35,0)

\ArrowLine(10,100)(60,100)
\ArrowLine(10,150)(60,150)
\ArrowLine(10,100)(10,150)
\ArrowLine(60,150)(60,100)
\ArrowLine(60,100)(35,70)
\ArrowLine(35,70)(10,100)

\ArrowLine(35,50)(35,70)
\ArrowLine(10,150)(10,170)
\ArrowLine(60,170)(60,150)

\Text(-10,135)[]{$k_2+q_1$}
\Text(82,135)[]{$k_2+q_2$}
\Text(35,160)[]{$k_2$}
\Text(35,45)[]{$q_1-q_2$}
\Text(35,25)[]{$generic$}
\Text(5,85)[]{$k_1+q_1$}
\Text(70,85)[]{$k_1+q_2$}
\Text(35,107)[]{$k_1-k_2$}
\Text(10,175)[]{$q_1$}
\Text(60,175)[]{$q_2$}

\Text(  0,118)[]{$m_3$}
\Text( 25, 70)[]{$m_1$}
\Text( 73,118)[]{$m_4$}
\Text( 50, 70)[]{$m_2$}
\Text( 35,140)[]{$m_5$}
\Text( 35, 87)[]{$m_6$}
\Vertex(10,100){2}
\Vertex(60,100){2}
\Vertex(10,150){2}
\Vertex(60,150){2}
\Vertex(35,70){2}

\DashLine(120,150)(170,150){3}
\Line(120,100)(170,100)
\DashLine(120,100)(120,150){3}
\DashLine(170,150)(170,100){3}
\Line(145,70)(120,100)
\Line(170,100)(145,70)


\Line(145,50)(145,70)
\DashLine(120,150)(120,170){3}
\DashLine(170,170)(170,150){3}

\Vertex(120,100){2}
\Vertex(170,150){2}
\Vertex(145,70){2}
\Vertex(120,150){2}
\Vertex(170,100){2}

\Text(145,25)[]{$Case\; 7$}


\DashLine(230,100)(230,150){3}
\DashLine(280,150)(280,100){3}
\DashLine(255,70)(230,100){3}
\DashLine(280,100)(255,70){3}


\DashLine(230,150)(230,170){3}
\DashLine(280,170)(280,150){3}
\Line(255,50)(255,70)
\DashLine(230,150)(280,150){3}
\Line(230,100)(280,100)

\Vertex(230,100){2}
\Vertex(230,150){2}
\Vertex(280,100){2}
\Vertex(280,150){2}
\Vertex(255,70){2}


\Text(255,25)[]{$Case\; 8$}


\DashLine(340,100)(340,150){3}
\DashLine(390,150)(390,100){3}
\DashLine(365,70)(340,100){3}
\DashLine(390,100)(365,70){3}


\DashLine(340,150)(340,170){3}
\DashLine(390,170)(390,150){3}
\Line(365,50)(365,70)
\DashLine(340,150)(390,150){3}
\DashLine(340,100)(390,100){3}

\Vertex(340,100){2}
\Vertex(340,150){2}
\Vertex(390,100){2}
\Vertex(390,150){2}
\Vertex(365,70){2}


\Text(365,25)[]{$Case\; 10$}

\vspace{-10pt}

\end{picture}
\end{figure}


\vspace{1.cm}
\end{document}